\documentclass[fleqn,10pt]{wlscirep}
\usepackage[utf8]{inputenc}
\usepackage[T1]{fontenc}
\usepackage{mathrsfs}
\usepackage{amsmath,amsthm,amssymb}
\usepackage[normalem]{ulem}
\useunder{\uline}{\ul}{}

\title{Community-Centers Identify Robust Biomarkers in High-Dimensional, Low-Sample-Size Gene Expression Data}

\author[1,2,8]{Ruiqi Li} 
\author[2]{Te Bai} 
\author[3,*]{Renaud Lambiotte} 
\author[4,*]{Orr Levy} 
\author[5]{Paul Expert} 
\author[2,6,*]{Daqing Li} 
\author[7]{Shlomo Havlin} 

\affil[1]{Beihang Valencia Polytechnic Institute, Beihang University, Hangzhou 311115, China}
\affil[2]{Hangzhou International Innovation Institute, Beihang University, Hangzhou 311115, China}
\affil[3]{Mathematical Institute, University of Oxford, Oxford OX2 6GG, UK}
\affil[4]{Faculty of Engineering, Bar-Ilan University, Ramat Gan 5290002, Israel}
\affil[5]{Global Business School for Health, Faculty of Population Health Sciences, University College London, London WC1E 6BT, UK}
\affil[6]{College of Safety Science and Engineering, Civil Aviation University of China, Tianjin 300300, China}  
\affil[7]{Department of Physics, Bar-Ilan University, Ramat Gan 5290002, Israel}
\affil[8]{Shanghai Key Laboratory of Intelligent Information Processing, Fudan University, Shanghai 200433, China}

\affil[*]{Corresponding authors. renaud.lambiotte@maths.ox.ac.uk, levyorr@biu.ac.il, daqingl@buaa.edu.cn}


\begin{abstract}
High-dimensional, low-sample-size bulk gene expression data poses a fundamental challenge in transcriptomics, which typically includes tens of thousands of genes but relatively few samples, leading to overfitting and unstable selection for key features as biomarkers.
We propose a regression by community centers (RCC) framework tailored for such data. RCC converts gene expression data into a feature proximity network and leverages the phase transition of the giant connected component to identify a critical distance threshold that yields a sparse yet maximally informative network representation, indicated by a minimum normalized shortest compression length. Intuitively, at the criticality, the network balances fragmentation and over-connectivity, allowing meaningful gene-gene associations communities to emerge while filtering noises, 
such that community centers capture the most informative and non-redundant signals. 
These representative genes are then fed to a simple ordinary least squares (OLS) model for downstream predictions. 
Across five bulk gene expression datasets, RCC displays exceptional prediction performance, clearly outperforming benchmark methods, remains robust to missing data and noises, and does not rely on external biological knowledge. These results suggest that network-based representations provide an effective and general framework for high-dimensional, low-sample-size prediction tasks beyond transcriptomics and illustrate how network-science ideas can support robust learning in data-scarce regimes.

\end{abstract}
\begin{document}

\flushbottom
\maketitle

\section*{Introduction}
Bulk gene expression data is crucial for biomedical studies, including identifying disease-relevant genes and building reliable predictive models of clinical outcomes \cite{li2008network}. 
However, bulk gene expression datasets often contain tens of thousands of measured genes (features), with orders of magnitude smaller number of patient samples \cite{gao2015high}. This can be due to high collection costs and ethical or practical constraints, and such a commonly encountered high-dimensional and low-sample-size setting \cite{corpas2025bridging,wang2017genome}, i.e., the dimensionality $d$ far exceeds the sample size $n$ ($d \gg n$), pose great challenges for machine learning models \cite{liu2017deep,bishop2006pattern} to achieve robust and strong predictive performance \cite{hastie2009elements, ruiz2023high}. 
This poses a fundamental statistical challenge: as data are so sparse relative to the feature space, models cannot cope with the excess of degrees of freedom. The likelihood of spurious correlations driving results increases, increasing the risk of overfitting \cite{hastie2009elements, ruiz2023high}.

To mitigate the challenges posed by high dimensionality, a common strategy is dimensionality reduction\cite{liu2017deep,van2009dimensionality}. Principal component analysis (PCA) \cite{abdi2010principal} and related methods project data onto a lower-dimensional subspace that captures maximal variance. While reducing dimensionality can improve downstream predictive performance, such approaches still require measuring all original features as input and therefore do not reduce the cost of acquiring high-dimensional data.
In addition, PCA is based on linear transformations, an assumption that is frequently violated in biological systems, where regulatory interactions and molecular responses are inherently nonlinear \cite{islam2017comparing}.
Consequently, linear dimensionality reduction may fail to recover critical structure in biological data.
An alternative is sparse modeling via feature selection\cite{guyon2003introduction,fan2007statistical}. The least absolute shrinkage and selection operator (LASSO) \cite{tibshirani1996regression} addresses this by augmenting ordinary least squares (OLS) regression (see Methods) with an $\ell_1$ penalty, favouring sparse non-zero coefficients thereby selecting a subset of predictive variables. However, when features are highly correlated or redundant, $\ell_1$ regularization tends to select one representative feature from a correlated group, which can yield unstable, non-reproducible models and reduce scientific interpretability \cite{su2017false}. More broadly, regularization can shrink coefficients or enforce sparsity \cite{tibshirani1996regression}, but it does not resolve the underlying data poverty in the $d \gg n$ regime, which in turn complicates the choice of the regularization penalty $\lambda$.
To incorporate additional structure, Network-Constrained LASSO (NC-LASSO) \cite{li2008network} leverages prior knowledge from external resources, such as gene regulatory networks, to guide variable selection. In addition to the $\ell_1$ sparsity term, it introduces a graph Laplacian penalty that smooths regression coefficients over the network, reflecting the assumption that genes connected in the regulatory graph should have similar effects. Nevertheless, LASSO and its variants \cite{grosenick2013interpretable,li2008network,hallac2015network} can remain sensitive to noise, potentially selecting spurious variables while discarding truly relevant features prematurely \cite{su2017false}.

More recently, motivated by a principle similar to NC-LASSO, the state-of-the-art method PLATO \cite{ruiz2023high} leverages auxiliary biomedical knowledge encoded in a large heterogeneous knowledge graph.
This graph integrates multiple sources of domain information, including gene regulatory relationships, gene--function associations, and other curated biological relations, to alleviate data scarcity.
Using a trainable message-passing mechanism, PLATO learns vector representations of genes (the input features) and uses them to regularize a multilayer perceptron.
Although PLATO achieves state-of-the-art performance, it depends heavily on rich external knowledge resources, which may be unavailable in other data-scarce domains.

\begin{figure}[ht] \centering
\includegraphics[width=\linewidth]{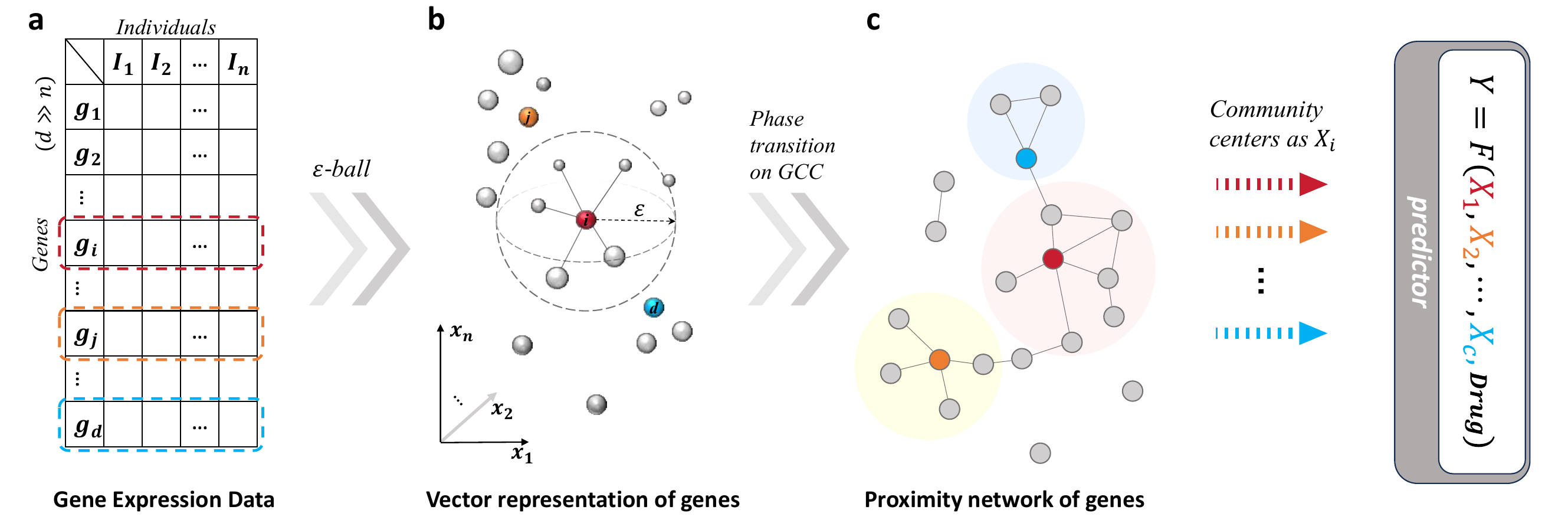}
\caption{\textbf{The framework of regression by community center (RCC) algorithm.} 
\textbf{a} The original high-dimensional, low-sample-size gene expression data, with the dimension $d$ much larger than the number of samples $n$. Here, for a better visualization, we transpose the data with each row denoting a gene, i.e., a feature, and each column denote a sample, i.e., an individual. 
\textbf{b} Vector representation of genes and illustration for $\varepsilon$-ball method. With the gene expression values from samples, we project $d$ features into a (small) $n$-dimensional Euclidean space. We treat each feature as a node, and if another node falls within the range of $\varepsilon$ of the focal node, then there will be an edge between them. After traversing every node, we can obtain a proximity network. 
Here, the choice of $\varepsilon$ is guided by its value at the phase transition of GCC (Fig. \ref{fig.constructingNetwork}a). 
\textbf{c} Constructed proximity network of genes and identified community centers, which are input of the predictor. 
On the constructed proximity network, we apply community detection algorithm to obtain communities and their centers, we then feed them to a predictor for downstream tasks. 
}
\label{fig.framework}
\end{figure}

In this work, we instead derive structural information directly from the observed data, without relying on external resources. We treat each feature, e.g., a gene, as a node and represent it by its values, e.g., expression value, across samples, i.e., a vector in $\mathbb{R}^n$ (Fig.~\ref{fig.framework}a). We then construct a feature proximity network using the $\varepsilon$-ball rule: two features are connected if the distance between their sample-value vectors is less than a threshold $\varepsilon$ (Fig.~\ref{fig.framework}b). When $\varepsilon$ is very small, the resulting network consists of mostly isolated nodes; when $\varepsilon$ exceeds the diameter of the feature cloud, the network becomes a complete graph. Between these extremes, we observe a percolation phase transition with the emergence of a giant connected component (GCC) and a set of smaller, disconnected components at a critical value of the threshold $\varepsilon$. Strikingly, we find that the GCC of the proximity networks constructed at the critical threshold, $\varepsilon_c$, encode the largest amount of structural information, as reflected by a minimum in structural entropy \cite{Choi2012Compression,shannon1948entropy,shannon1953lattice}. 
We then turn our attention to the community structure of the proximity network obtained at the critical threshold. When a nontrivial community structure is present (Fig.~\ref{fig.framework}c and Fig. \ref{fig.constructingNetwork}d-e), which is characterised by a high modularity value\cite{newman2006modularity}, we hypothesise that the community centres \cite{shi2024local} are candidate features with high information content for prediction tasks.
We justify this assumption by observing that community centers have a dominant position in their neighborhood and are relatively far away from other more influential nodes \cite{shi2024local}, thus summarising a significant amount of local information.
Moreover, selecting one center per community also promotes diversity by reducing redundancy among similar features. Feeding these community-center features and drug variables to even a simple ordinary least squares (OLS) predictor, our method consistently outperform all benchmarks and achieve the best overall performance across the five transcriptomic datasets, without relying on external knowledge. We refer to this framework as \emph{regression by community centers} (RCC). 
Beyond predictive gains, our findings deepen the understanding of how high-dimensional vector data can be mapped onto informative network representations, and they suggest a principled, physics-inspired route for learning in data scarce environments.

\section*{Results}

\begin{table}[] \centering
\begin{tabular}{|l|c|c|c|c|c|}
\hline
Datasets             & BRCA   & PDAC   & MNSCLC & CM     & CRC   \\ \hline
\#features $d$          & 12,693 & 12,932 & 15,390 & 13,183 & 18,206  \\ \hline
\#patients $n$       & 41     & 38     & 27     & 32     & 51     \\ \hline
\#drugs applied $u$     & 18     & 15     & 19     & 12     & 21     \\ \hline
\#experiments on mice & 768    & 575    & 523    & 484    & 803 \\   \hline
\end{tabular}
\caption{\textbf{Basic dataset statistics.} Each dataset contains $d$ gene features, where each feature value represents the expression level of a gene measured from tumor tissue, and the number of patients $n$ is limited, with $n \ll d$. For each patient, the tumor tissue was expanded and implanted into mice, which were then assigned treatments from a set of $u$ options consisting of single drugs or drug combinations.
 BRCA, PDAC, MNSCLC, CM, and CRC stand for Breast Cancer, Pancreatic Ductal Adenocarcinoma, Metastatic Non-Small Cell Lung Cancer, Cutaneous Melanoma, and Colorectal Cancer, respectively.}\label{table.datasets}
\end{table}

\subsection*{Datasets}
To evaluate our proposed framework, we analyze five high-dimensional, low-sample-size transcriptomic datasets \cite{gao2015high} (Table~\ref{table.datasets}). These datasets consist of gene-expression measurements obtained from patients' tumor tissue and are preprocessed using log-transformation and normalization \cite{ruiz2023high} (Table~\ref{tab.rawData}). The response variable is the change in tumor volume following treatment, measured in mouse experiments (Table~\ref{tab.rawData}).
In the experiments, tumor tissue from each patient was processed to remove necrotic and stromal components. Approximately 20--30 mg of the resulting tissue fragment was then implanted subcutaneously into the flank of an athymic nude female mouse using a trocar. Successfully engrafted tumors were passaged (typically passages 4--10) to reduce genetic drift. Each mouse bearing a tumor was subsequently treated with a single drug or a drug combination. Tumor size was monitored, and the minimum change in tumor volumn within ten days after treatment is recorded as the outcome.
We illustrate for the metastatic non-small cell lung cancer (MNSCLC) dataset that includes expression measurements for 15{,}390 genes from only 27 patients. Tumor tissue from each patient is expanded to generate roughly 20 mouse implants, yielding a total of 523 vivo experiments. For each experiment, a specific drug or a drug combination is administered, and the minimum tumor volume change after ten days of treatment was recorded as the response variable (Table~\ref{tab.rawData}, last column). This response serves as the prediction target for machine learning models. For context, standard treatment schedules are typically 21 days and may extend to 90 days in the absence of disease progression or adverse events.

\begin{table}[] \centering  
\begin{tabular}{|cc|ccccc|cccc|c|}
\hline
\multicolumn{2}{|c|}{MNSCLC dataset} & \multicolumn{5}{c|}{Gene expression of subject}                                                                        & \multicolumn{4}{c|}{Drugs applied on mice}                                                        & Response \\ \hline
\multicolumn{1}{|c|}{Subject}      & Mouse     & \multicolumn{1}{c|}{$g_1$} & \multicolumn{1}{c|}{$g_2$} & \multicolumn{1}{c|}{$g_3$} & \multicolumn{1}{c|}{$\cdots$} & $g_n$ & \multicolumn{1}{c|}{$d_1$} & \multicolumn{1}{c|}{$d_2$} & \multicolumn{1}{c|}{$\cdots$} & $d_u$ & Y        \\ \hline
\multicolumn{1}{|c|}{1}               & 1                       & \multicolumn{1}{c|}{-1.87} & \multicolumn{1}{c|}{0.49}  & \multicolumn{1}{c|}{-0.45} & \multicolumn{1}{c|}{$\cdots$} & -0.36  & \multicolumn{1}{c|}{1}     & \multicolumn{1}{c|}{0}     & \multicolumn{1}{c|}{$\cdots$} & 0     & 1.27    \\ \hline
\multicolumn{1}{|c|}{1}               & 2                       & \multicolumn{1}{c|}{-1.87} & \multicolumn{1}{c|}{0.13}  & \multicolumn{1}{c|}{-0.45} & \multicolumn{1}{c|}{$\cdots$} & -0.36  & \multicolumn{1}{c|}{0}     & \multicolumn{1}{c|}{1}     & \multicolumn{1}{c|}{$\cdots$} & 0     & 1.32    \\ \hline
\multicolumn{1}{|c|}{2}               & 3                       & \multicolumn{1}{c|}{-0.15} & \multicolumn{1}{c|}{-0.01} & \multicolumn{1}{c|}{-0.13} & \multicolumn{1}{c|}{$\cdots$} & 1.05  & \multicolumn{1}{c|}{0}     & \multicolumn{1}{c|}{0}     & \multicolumn{1}{c|}{$\cdots$} & 1     & 0.64     \\ \hline
\multicolumn{1}{|c|}{3}               & 4                       & \multicolumn{1}{c|}{0.68}  & \multicolumn{1}{c|}{-2.1}  & \multicolumn{1}{c|}{-0.74} & \multicolumn{1}{c|}{$\cdots$} & -0.21 & \multicolumn{1}{c|}{1}     & \multicolumn{1}{c|}{0}     & \multicolumn{1}{c|}{$\cdots$} & 0     & -0.83    \\ \hline
\multicolumn{1}{|c|}{3}               & 5                       & \multicolumn{1}{c|}{0.68}  & \multicolumn{1}{c|}{-2.1}  & \multicolumn{1}{c|}{-0.74} & \multicolumn{1}{c|}{$\cdots$} & -0.21 & \multicolumn{1}{c|}{0}     & \multicolumn{1}{c|}{1}     & \multicolumn{1}{c|}{$\cdots$} & 1     & -1.42     \\ \hline
\multicolumn{1}{|c|}{$\cdots$}              & $\cdots$                      & \multicolumn{1}{c|}{$\cdots$}    & \multicolumn{1}{c|}{$\cdots$}    & \multicolumn{1}{c|}{$\cdots$}    & \multicolumn{1}{c|}{$\cdots$} & $\cdots$    & \multicolumn{1}{c|}{$\cdots$}    & \multicolumn{1}{c|}{$\cdots$}    & \multicolumn{1}{c|}{$\cdots$} & $\cdots$    & $\cdots$       \\ \hline
\multicolumn{1}{|c|}{27}              & 523                     & \multicolumn{1}{c|}{0.16}  & \multicolumn{1}{c|}{-0.3}  & \multicolumn{1}{c|}{0.63}  & \multicolumn{1}{c|}{$\cdots$} & 0.26  & \multicolumn{1}{c|}{0}     & \multicolumn{1}{c|}{0}     & \multicolumn{1}{c|}{$\cdots$} & 1     & 2.02      \\ \hline
\end{tabular}
\caption{\textbf{The format of datasets.} Here, we demonstrate the MNSCLC dataset as an example. Each row correspond to a record from an experiment on a mouse that is implanted with a tumor tissue from the subject after taking certain drug(s). The gene expression are measured from the tumor tissue of the subject. All datasets used in this work are of the same format.}\label{tab.rawData}
\end{table}

\subsection*{Constructing proximity network from high-dimensional tabular data}
To avoid the computational burden of constructing a full correlation matrix over $d$ features (genes), our method instead seeks a small set of representative features. 
We treat each feature as a node and use its values across samples as a vector representation. This embeds the $d$ nodes into an $n$-dimensional Euclidean space (Fig.~\ref{fig.framework}a-b), where typically $n \ll d$.
To infer proximity network connectivity, we adopt the $\varepsilon$-ball rule: two nodes are connected if the distance between their feature vectors is smaller than $\varepsilon$ (Fig.~\ref{fig.framework}b). Compared with $k$-nearest-neighbor graphs, the $\varepsilon$-ball construction avoids spurious long-range edges that can arise for peripheral points. Empirically, shorter distances in this embedding correspond to higher correlation (Supplementary Note~1 and Supplementary Fig.~S1) and are also associated with smaller $p$-values, indicating stronger statistical signals (Supplementary Note~1 and Supplementary Fig.~S2). Moreover, $\varepsilon$-ball graph construction can be accelerated using a modified R-tree \cite{guttman1984r,li2017simple}, which supports efficient spatial queries. 

\begin{figure}[!htb] \centering
\includegraphics[width=\linewidth]{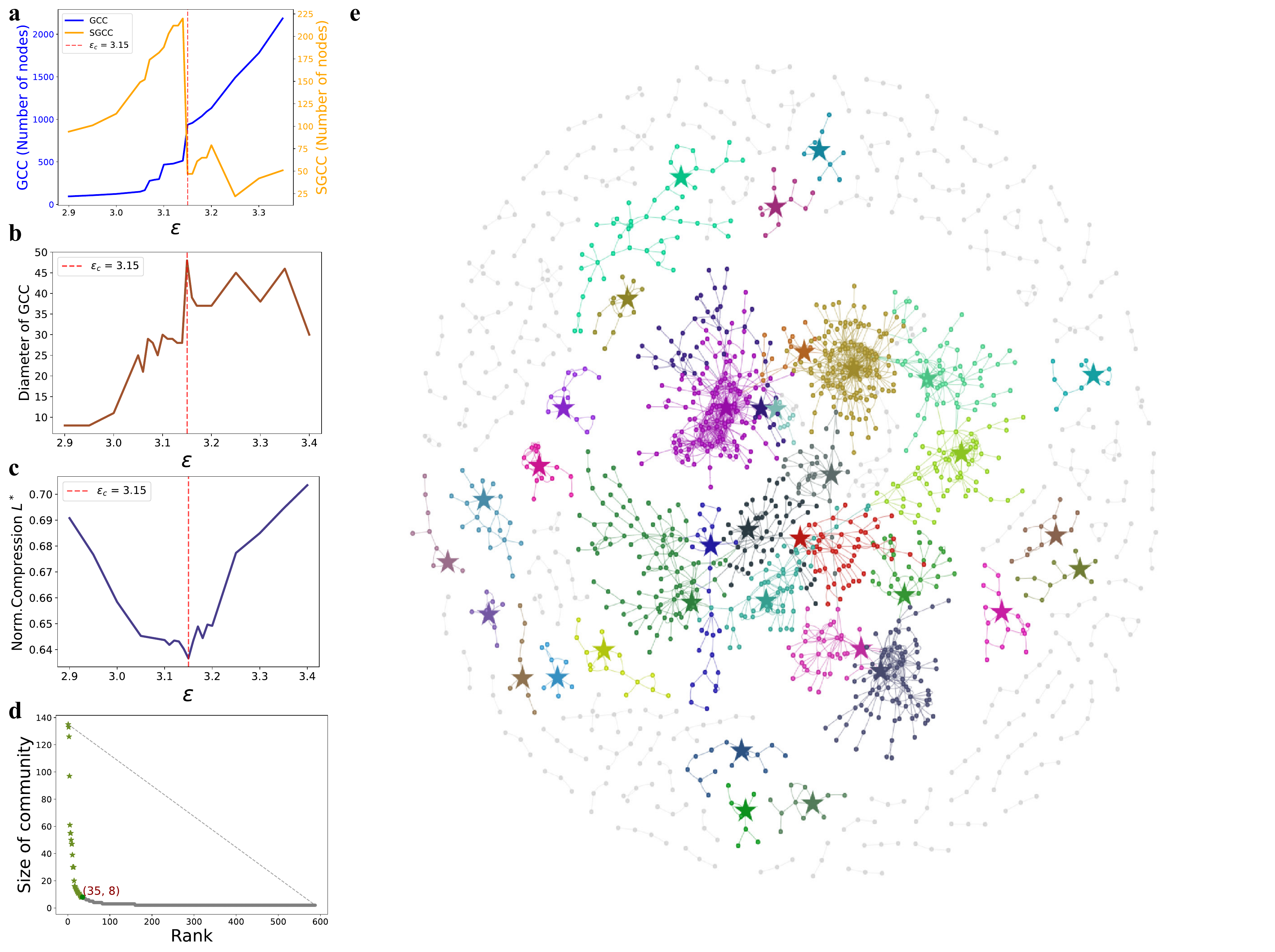}
\caption{\textbf{Proximity network construction and identification of community centers.}
\textbf{a} As $\varepsilon$ increases in the $\varepsilon$-ball construction, the giant connected component (GCC) increases and exhibits a phase transition, as indicated by the sharp peak of the second-largest connected component. The critical threshold $\varepsilon_c$ is marked by the vertical red dashed line.
\textbf{b} The diameter of GCC as a function of $\varepsilon$, which also peaks at $\varepsilon_c$. Here, shortest-path distances are computed only for nodes in the GCC. 
\textbf{c} Normalized shortest compression length of the constructed proximity networks (see Methods) versus $\varepsilon$, reaching a minimum near $\varepsilon_c$.
\textbf{d} Rank--size distribution of communities detected by the Louvain algorithm for the proximity network constructed at $\varepsilon_c$. Only communities ranked above the elbow point are retained for further prediction task. The elbow point is defined as the point with the largest perpendicular distance to the gray dashed line connecting the first and last points; its rank and community size are shown in parentheses. 
In this example, there are 35 communities above the elbow point, they reside in 20 finite components, with the GCC comprising 16 communities, and other 19 components each forming a community (Table \ref{table.modularity}). 
\textbf{e} Community visualization and the corresponding community centers. Colors indicate community assignments, and community centers are highlighted by stars. Nodes shown in light gray correspond to small communities below the elbow point in \textbf{d}.}
\label{fig.constructingNetwork}
\end{figure}

Choosing an appropriate threshold $\varepsilon$ is nontrivial. When $\varepsilon$ is very small, the resulting graph contains mostly isolated nodes; when $\varepsilon$ exceeds the diameter of the feature cloud, the graph becomes (nearly) complete, which is again uninformative. As $\varepsilon$ increases, we observe a phase transition marked by the emergence of a giant connected component (GCC) \cite{bunde1996fractals}. Notably, at the critical point (Fig.~\ref{fig.constructingNetwork}a; Supplementary Fig.~S3 for other datasets), the diameter of the GCC reaches its maximum (Fig.~\ref{fig.constructingNetwork}b; Supplementary Fig.~S4), while the normalized shortest compression length \cite{Choi2012Compression,sun2020revealing} (see Methods) reaches its minimum (Fig.~\ref{fig.constructingNetwork}c; Supplementary Fig.~S5). This minimum indicates maximal structural regularity and, correspondingly, maximal information encoded in the network topology \cite{Choi2012Compression,sun2020revealing}. 
We therefore convert the high-dimensional continuous feature vectors into a discrete network representation (Fig.~\ref{fig.framework}c; Fig.\ref{fig.constructingNetwork}e) constructed at the critical threshold $\varepsilon_c$.


\subsection*{Community centers as key features}

Complex networks often exhibit community structure, characterised by dense connections among nodes within the same community and sparser connections across communities \cite{girvan2002community,fortunato202220,shi2024local}. The proximity networks constructed from high-dimensional vector data indeed display nontrivial community structure(Fig.~\ref{fig.constructingNetwork}e), quantified by a high modularity \cite{newman2006modularity,blondel2008fast} (Table \ref{table.modularity} and Supplementary Table~S1). Because edges in our graphs reflect strong similarity and empirically correspond to high correlation (see details in Supplementary Note~1 and Supplementary Figs.~1-2), nodes within the same community tend to be more correlated and more alike than nodes in different communities. This suggests that a representative node per community can be used to represent a group of similar features. Selecting one representative from each community also promotes diversity, since different communities capture distinct aspects of the original feature space. We therefore treat community centers as candidate key features of the original high-dimensional data.

We define a community center as a node that is both highly influential and well separated from other more influential nodes \cite{shi2024local}. Influence can be assessed using standard centrality measures \cite{shi2024local,guo2026improved} or other metadata reflecting node importance, and separation is measured by the distance to nodes with higher influence. Community centers can be identified jointly with community detection, e.g., via the local search (LS) algorithm \cite{shi2024local}, or determined post hoc after obtaining a partition using other community detection methods, e.g., Louvain algorithm \cite{blondel2008fast} (see Methods). For instance, given the communities detected by the Louvain algorithm \cite{blondel2008fast}, a simple post hoc choice is to select, within each community, the node with the highest value on a centrality measure such as coreness \cite{kitsak2010identification,li2024evolution}, $h$-index \cite{lu2016h}, degree, closeness \cite{evans2022linking}, or a weighted combination thereof (see Methods).

As the proximity network constructed at the critical threshold $\varepsilon_c$ can comprise several disconnected components, we apply community detection algorithm to each component. The GCC generally comprises tens of communities, while other small components generally each form a single community (Table \ref{table.modularity} and Supplementary Table S1). The resulting rank--size distributions of communities is quite heterogeneous with many small communities in the tail part (Fig.~\ref{fig.constructingNetwork}d). An overabundance of such small communities can yield too many community centers, potentially degrading the performance of downstream predictors. To improve robustness, we retain only centers from sufficiently large communities--specifically, those above the elbow point of the rank--size distribution of communities (Fig.~\ref{fig.constructingNetwork}d,e; Supplementary Fig.~S6 and Supplementary Tables~S2 for other datasets).

\begin{table}[!htbp] \centering
\begin{tabular}{|l|c|c|c|c|c|}
\hline
Datasets  & BRCA      & PDAC        & MNSCLC            & CM         & CRC          
\\ \hline
\#nodes $N$              & 2,781          & 2,078          & 2,502          & 2,230          & 2,186          \\ \hline
\#edges $E$              & 3,769          & 2,843          & 3,087          & 4,466          & 4,832          \\ \hline 
\#connected components      & 15          & 18      & 20          & 9          & 6     \\ \hline
\#communities in GCC        & 17          & 16      & 16          & 17          & 20     \\ \hline
\#communities in smaller components        & 14          & 17      & 19          & 8          & 5     \\ \hline
Q           & 0.9312          & 0.9248      & 0.9276          & 0.7574          & 0.7870     \\ \hline
\end{tabular}
\caption{\textbf{Basic statistics of the constructed proximity networks at criticality.} 
The number of communities refers to the ones above the elbow point based on rank-size distribution of community partitions from Louvain algorithm \cite{blondel2008fast} (Fig. \ref{fig.constructingNetwork}d; Supplementary Fig.~S6). Communities above the elbow point correspond to relatively large ones. 
In all five datasets, the GCC comprises multiple communities, and other communities each corresponds to a connected component. Compared to Louvain method, LS algorithm \cite{shi2024local} identifies fewer large communities above the elbow point (Supplementary Fig.~S7; Supplementary Table~S1). Q refers to the modularity score \cite{newman2006modularity}. 
} \label{table.modularity}
\end{table} 

\subsection*{Prediction performance}
To demonstrate the effectiveness of our framework, termed \emph{regression by community centers} (RCC), we deliberately adopt the simplest predictor: ordinary least squares (OLS) (Fig.~\ref{fig.framework}c). Specifically, we use the features corresponding to the identified community centers together with drug-related variables as inputs to the OLS model to predict tumor-volume change.
We benchmark the RCC against 11 widely used and diverse baselines, including regularized linear regression (Ridge)\cite{hoerl1970ridge}, feature-selection methods (LASSO \cite{tibshirani1996regression}, GraphNet \cite{grosenick2013interpretable}, and NC-LASSO \cite{li2008network}), dimensionality reduction (PCA \cite{abdi2010principal}), a tree-based ensemble (XGBoost \cite{chen2016XGBoost}), k-means\cite{lloyd1982least} (which suffers from choosing an appropriate number of clusters, Supplementary Fig. S8), and several tabular deep-learning models (MLP\cite{lecun2015deep,ng2004feature}, TabTransformer\cite{huang2020tabtransformer}, and TabNet\cite{arik2021tabnet}). We also compare the RCC to the current state-of-the-art method PLATO, which leverages a large auxiliary heterogeneous knowledge graph to learn feature embeddings that regularize an MLP \cite{ruiz2023high}.
Remarkably, despite using a simple OLS predictor, RCC consistently outperforms all baselines across the five transcriptomic datasets (Table~\ref{table.datasets}) and achieves the best overall performance (Table~\ref{table.performance}). 
If we change the community detection method in RCC from using Louvain algorithm \cite{blondel2008fast} to other methods, e.g. the LS algorithm \cite{shi2024local}, which does not optimise modularity directly but rather directly identifies centres, the RCC still achieves better perfoemance good performance, only slightly under-performs than PLATO \cite{ruiz2023high} on BRCA dataset (Table~\ref{table.performance}). This also highlights the potential of utilising network-science principles for robust learning in data-scarce regimes.   
If we change the community detection method in RCC from using Louvain algorithm \cite{blondel2008fast} to other methods, e.g. the LS algorithm \cite{shi2024local}, which does not optimise modularity directly but rather directly identifies centres, the RCC still achieves a higher prediction than other methods, only closely beaten by PLATO \cite{ruiz2023high} on the BRCA dataset (Table~\ref{table.performance}). This also highlights the potential of utilising network-science principles for robust learning in data-scarce regimes.   
In addition, we expect that replacing OLS with a more expressive predictor could further improve performance within the RCC framework. 
Importantly, the performance of the RCC framework degrades when applied to proximity networks constructed away from criticality (Supplementary Table~S3). This observation confirms the intuition that proximity network at criticality, which avoids both sparsity and redundancy, possess a unique informational expressivity.

\begin{table}[!hb] \centering
\begin{tabular}{|l|c|c|c|c|c|}
\hline
Datasets               & BRCA                 & PDAC                 & MNSCLC               & CM                   & CRC                  \\ \hline
PCA                    & 0.452$\pm$0.000          & 0.232$\pm$0.121          & 0.156$\pm$0.113          & 0.070$\pm$0.000          & 0.193$\pm$0.163          \\ \hline
Ridge                  & 0.538$\pm$0.000          & 0.344$\pm$0.000          & 0.153$\pm$0.000          & 0.390$\pm$0.000          & 0.376$\pm$0.000          \\ \hline
LASSO                  & 0.470$\pm$0.000          & 0.346$\pm$0.000          & 0.168$\pm$0.000          & 0.431$\pm$0.000          & 0.400$\pm$0.000          \\ \hline
XGBoost                & 0.347$\pm$0.000          & 0.363$\pm$0.000          & -0.02$\pm$0.000          & 0.225$\pm$0.000          & 0.354$\pm$0.000          \\ \hline
k-means         & 0.400$\pm$0.054          & 0.233$\pm$0.065          & 0.574$\pm$0.032          & 0.284$\pm$0.074          & 0.333$\pm$0.055          \\ \hline
GraphNet               & 0.350$\pm$0.069          & 0.249$\pm$0.018          & 0.169$\pm$0.030          & 0.277$\pm$0.099          & 0.125$\pm$0.061          \\ \hline
NC LASSO               & 0.458$\pm$0.083          & 0.327$\pm$0.053          & 0.210$\pm$0.014          & 0.339$\pm$0.044          & 0.220$\pm$0.030          \\ \hline
MLP                    & 0.240$\pm$0.067          & 0.289$\pm$0.047          & 0.128$\pm$0.126          & 0.322$\pm$0.043          & 0.355$\pm$0.022          \\ \hline
TabNet                 & 0.055$\pm$0.037          & 0.088$\pm$0.037          & 0.085$\pm$0.028          & 0.010$\pm$0.068          & 0.018$\pm$0.016          \\ \hline
TabTransformer         & 0.202$\pm$0.000          & 0.029$\pm$0.000          & 0.265$\pm$0.000          & 0.072$\pm$0.000          & 0.238$\pm$0.000          \\ \hline
{\ul PLATO}            & {\ul 0.583$\pm$0.019}     & 0.400$\pm$0.021          & 0.272$\pm$0.130          & 0.435$\pm$0.022          & 0.401$\pm$0.019          \\ \hline
\textbf{RCC (Louvain)} & \textbf{0.591$\pm$0.039}  & \textbf{0.504$\pm$0.047} & \textbf{0.665$\pm$0.047}    & \textbf{0.510$\pm$0.052} & \textbf{0.560$\pm$0.050} \\ \hline
\textbf{RCC (LS)}      & 0.573$\pm$0.038           & {\ul 0.500$\pm$0.051}    & {\ul 0.664$\pm$0.053}      & {\ul 0.464$\pm$0.049}    & {\ul 0.485$\pm$0.044}
\\ \hline
\end{tabular}
\caption{\textbf{Performance of RCC against 11 popular baselines across datasets.} RCC (Louvain) refers to using Louvain algorithm \cite{blondel2008fast} for detecting communities, and RCC (LS) utilizes LS \cite{shi2024local} algorithm for community partitioning. } \label{table.performance}
\end{table}

\subsection*{Robustness analysis}
Beyond strong predictive performance, the proposed RCC framework is robust to missing values and additive noise in the original vector data.

\begin{figure}[ht]\centering
\includegraphics[width=0.6\linewidth]{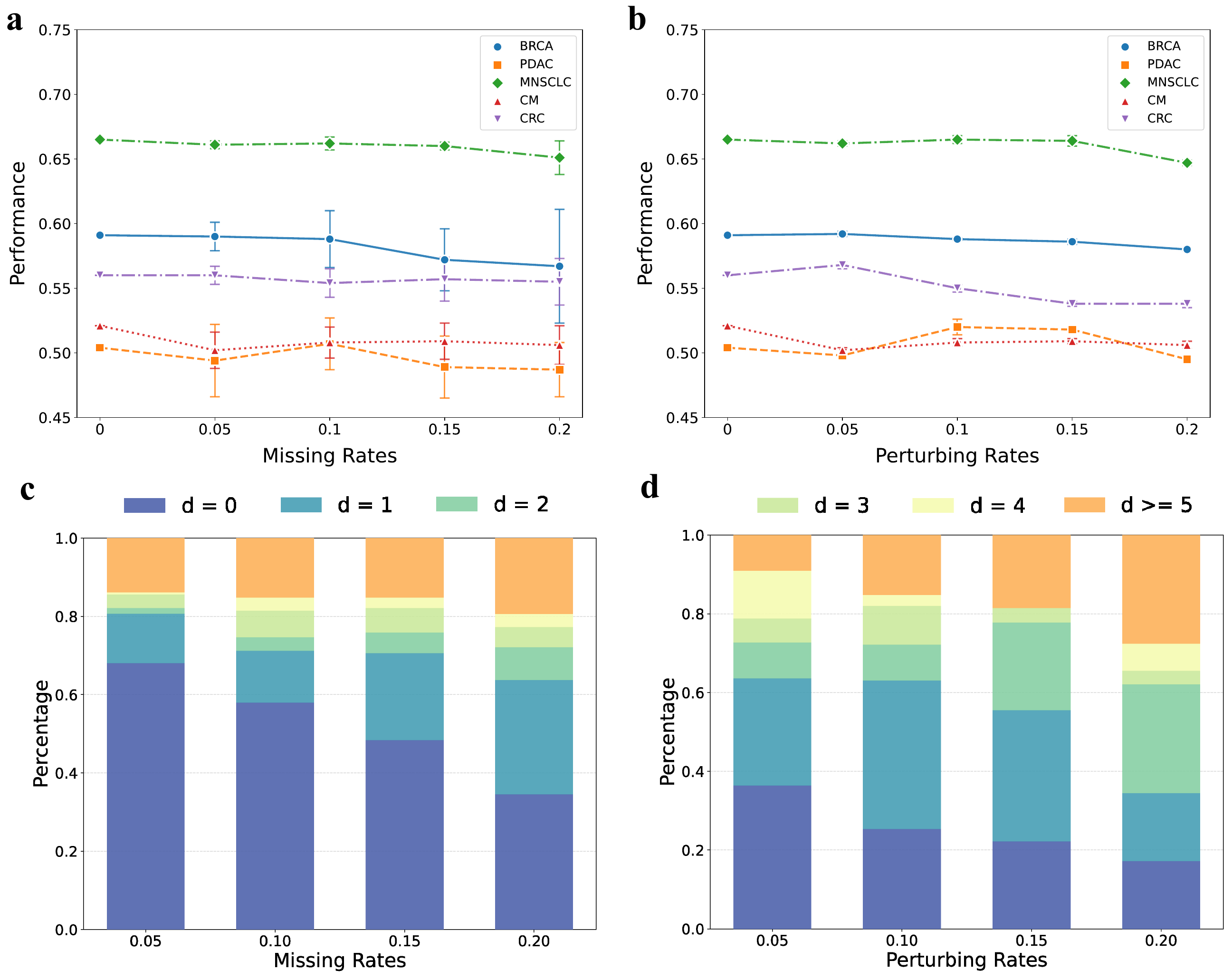}
\caption{\textbf{Robustness analysis.} The predictive performance of our RCC method after \textbf{a} randomly removing a certain fraction of features or \textbf{b} randomly perturbing a certain fraction of original gene expression value remain relatively high, which indicate that our method is quite robust to missing or noisy data. Taking the MNSCLC dataset as an example, we analyze the shortest path distance between the originally detected community centers and newly detected community centers after \textbf{c} randomly removing a certain fraction of features or \textbf{d} randomly perturbing a certain fraction of original gene expression value. Most of newly detected community centers either overlap with previous ones, i.e., shortest path length $d_{ij}$ between them is 0, or generally not too far from previous ones in the original constructed proximity network. The results reported here are the average from five random experiments. }
\label{fig.robustness}
\end{figure}

\paragraph{Robustness to missing features.}
We first evaluate RCC under missing-feature scenarios by randomly removing a fraction of features (i.e., columns) from the original datasets. It is found that RCC maintains its predictive performance even with substantial feature removal (Fig.~\ref{fig.robustness}a), so remains highly competitive in noisy environments. A plausible explanation is that the representative features identified by RCC are stable under such perturbations. For instance, after removing 20\% of the features, more than 60\% of the selected key features either coincide with those selected previously (shortest-path distance $d_{ij}=0$) or are directly adjacent to them in the constructed proximity network ($d_{ij}=1$) (Fig.~\ref{fig.robustness}c; Supplementary Fig.~S9 for other datasets).

\paragraph{Robustness to noisy measurements.}
We further assess stability under measurement noise by adding Gaussian perturbations to a fraction of the gene-expression entries. Specifically, we randomly select a given fraction of data items and add noise drawn from $\mathcal{N}(0,1)$. Even when 20\% of expression values are perturbed, RCC continues to achieve high predictive performance (Fig.~\ref{fig.robustness}b; Supplementary Fig.~S10 for other datasets). 

\begin{figure}[ht]\centering
\includegraphics[width=\linewidth]{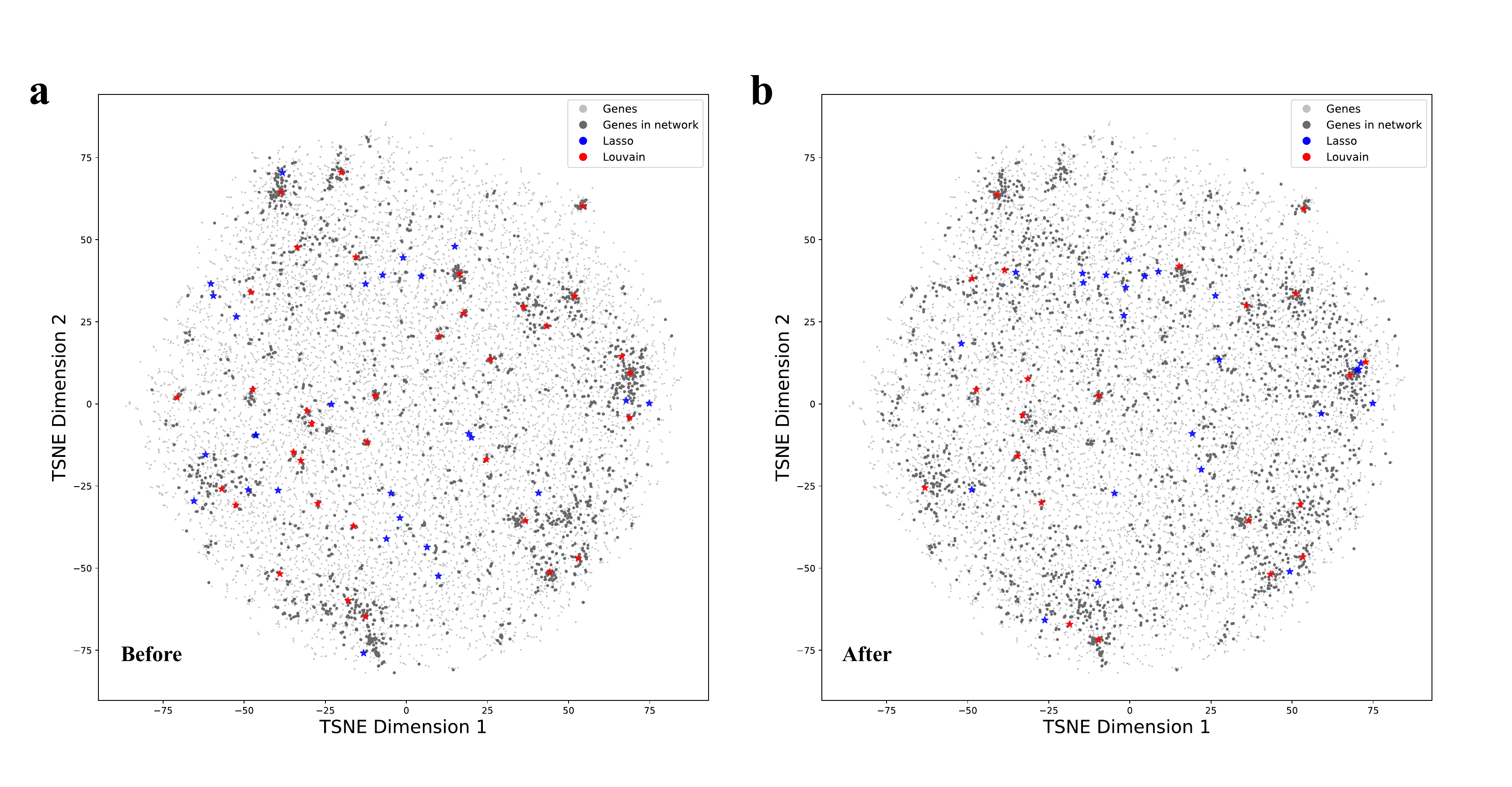}
\caption{\textbf{Key features selected by RCC algorithm and LASSO before and after perturbation.} \textbf{a} Key features selected by RCC algorithm and LASSO for the original data and \textbf{b} perturbed data that add a Gaussian noise $\mathcal{N}(0,1)$ to 20\% of entries. 
Here, the key features selected by RCC is highlighted in red, and the ones by LASSO are in blue. The key features selected by our RCC method is generally of a high density, while the ones selected by LASSO can be of quite low density. Each gray point correspond to a feature in the original gene expression data, and the dark gray ones are those appearing in the constructed proximity network by our method. 
We can clearly observe that the key features selected by LASSO change a lot, while the ones by RCC is relatively stable. 
Here, for better visualization, we keep the layout of the data after perturbation as the same with the one before perturbation. 
}
\label{fig.keyFeatures-RCC-lasso}
\end{figure}

\paragraph{Qualitative comparison of selected features.}
For a more intuitive comparison, we visualize the feature vectors using t-SNE \cite{maaten2008visualizing} and highlight the key features chosen by RCC and by LASSO. As shown in Fig.~\ref{fig.keyFeatures-RCC-lasso}, after perturbing 20\% of gene-expression values, the community centers selected by RCC remain close to those selected before perturbation, whereas LASSO may shift to features that are far away in the vector space (Supplementary Fig. S11). Moreover, RCC tends to select features located in denser regions of the vector space, while LASSO-selected features can be comparatively isolated (Fig.~\ref{fig.keyFeatures-RCC-lasso}; Supplementary Fig.~S12 for their density distribution).

 

\section*{Discussion}

High-dimensional, low-sample-size data remain challenging for machine learning models, because sparsity and overparameterization increase the risk of overfitting\cite{aggarwal2001surprising}, and datasets often contain additional complications such as measurement noise, missing values, and redundant (highly correlated) features. In this work, we propose the \emph{regression by community centers} (RCC) method, a fully data-driven framework that converts the original feature vectors into an informative network representation and selects representative features from the proximity network. Concretely, we treat each feature as a node and its sample values as an $n$-dimensional vector, then construct a proximity network using the $\varepsilon$-ball rule. A key step is choosing the distance threshold. Rather than tuning $\varepsilon$ heuristically, we use here the phase transition of the giant connected component (GCC) in the network to identify a principled critical value $\varepsilon_c$. At this critical point, the diameter of the largest connected component peaks while the normalized compression length reaches a minimum, indicating that the network topology is maximally informative and least redundant \cite{Choi2012Compression,sun2020revealing}, which avoids both noise-dominated sparsity and redundancy-dominated saturation. 
We find that if we apply our RCC framework to proximity networks constructed away from criticality, the performance degrades (Supplementary Table~S3), which is probably due to sparsity, above the critical threshold, or redundancy, below the critical threshold, in the constructed network. 
The resulting graph exhibits clear community structure, where many small communities are isolated connected components, those small ones below the elbow point of rank-size distribution of communities (Fig. \ref{fig.constructingNetwork}d) are excluded from further analysis. We then select the center of relatively large communities above the elbow point as representative key features. Strikingly, feeding only these community-center features (together with drug variables) into even a simple OLS predictor yields performance that consistently surpasses a broad range of baselines, including regularized linear models, network-regularized variants, and modern tabular deep-learning methods—without relying on external biological resources such as gene regulatory networks or heterogeneous knowledge graphs.

The effectiveness of the RCC can be understood through three complementary mechanisms.
First, proximity network construction acts as an adaptive filtering step: only sufficiently strong relationships are retained as edges, while small variations below the critical threshold are suppressed, thus, reducing sensitivity to noise.
Second, community centers provide a compact yet diverse feature set: because communities correspond to groups of similar (redundant) features, selecting one representative per community reduces collinearity while preserving coverage of distinct signals as communities cover the whole dataset. In addition, by looking up the description of those community center genes, many of them have been proved to be related to cancers or critical for important biochemical processes (Supplementary Tables S4-8). 
Third, community structure and center identification are intrinsically robust to moderate perturbations, helping RCC remain stable under missing-feature scenarios and measurement noise in the original gene-expression data.

Several extensions could further improve RCC and merit future investigation.
Different community-detection algorithms should also be examined, consistent with the no-free-lunch perspective for community detection \cite{peel2017ground}.
Beyond selecting only community centers, one could include a small neighborhood around each center as a fallback when a center feature is unavailable, or aggregate information from sampled neighbors to form community-level embeddings in a GraphSAGE-like manner \cite{hamilton2017inductive}.
The selected features set could then be further refined using forward, backward, or stepwise regression.
Given the success of PLATO \cite{ruiz2023high}, a natural direction is to augment features fed to RCC with auxiliary biomedical knowledge when such resources are available, while preserving a purely data-driven fallback when they are not. 
Finally, RCC is compatible with predictors more expressive than OLS, which may further improve predictive performance while retaining RCC's data-driven feature selection and robustness.

\section*{Methods}

\subsection*{R-tree for proximity network construction}
R-tree\cite{guttman1984r} is a balanced spatial searching tree, whose depth is proportional to the logarithmic of the number of nodes in the tree. 
Its core principle is to use a minimum bounding rectangle (MBR) to organize spatially nearby data points as a node in the searching tree, where each non-leaf internal node encloses all its child nodes. 
For a certain spatial query, the algorithm first traverse every entry of the root node of R-tree to determine whether the MBR of an entry intersects with the query region: if not, the whole brach connected to this entry can be skipped, which can significantly reduce unnecessary distance computations; if so, further go down the tree and repeat the process. 
Such a balanced tree can greatly accelerate searching for spatial objects \cite{guttman1984r} and thus the construction for the proximity network. 
Here, in our study, each feature $i$ is projected as a point in a $n$-dimensional vector space $(x^1_i,x^2_i,...,x^n_i)$, and then a leaf node in the R-tree with its MBR $(x^1_{min},x^1_{max},x^2_{min},x^2_{max},...,x^n_{min},x^n_{max})$ equals $(x^1_i,x^1_i,x^2_i,x^2_i,...,x^n_i,x^n_i)$. 
The R-tree construction process involves inserting $d$ leaf nodes, which requires further splitting and merging some internals nodes once they overflow or underflow. Simply put, the depth of the tree is $\log(d)$, each insertion at most require changing internal nodes on the path from it to the root node. As each entity is represented with a $2n$-dimensional vector, the determination of intersection is O($n$). Thus, the time complexity for R-tree construction is O($dn\log d$).  
The proximity network construction process depends on spatial query with each node as the center of an square whose BMR equals $(x^1_i-r/2,x^1_i+r/2,x^2_i-r/2,x^2_i+r/2,...,x^n_i-r/2,x^n_i+r/2)$. 
For leaf nodes that got intersected with such a BMR, we further compute their distance to the focal node, and establish an edge between two entities only if their distance is shorter than $r$. 
As the time complexity for each search is O(log $d$), the proximity network construction process, which requires $d$ searches in total, is proportional to O($dn\log d$). 
With R-tree structure, we can greatly speed up the proximity network construction process by avoiding calculating distance from the focal node to all other nodes. 

\subsection*{Identification of community centers}
Community centers are nodes with both a dominant position in its neighborhood and beyond and they are far away from other more influential nodes \cite{shi2024local}. In this sense, the community center can form its own attraction basin. 

The Local Search (LS) algorithm proposed by Shi et al. can directly identify community centers, which has a high influence $k$ and a long distance on network $l$ to another influential node, in a linear time complexity. LS first calculate the influence of each node, which can be approximated by degree or other centrality measures. Second, LS traverse each node $u$ and point $u$ to any adjacent node $v$ with $k_v\geq k_u$ and $k_v= \max\{k_z|z\in\mathbf{V}(u)\}$ (i.e., $v$ has the largest degree in the neighborhood of $u$). After traversing all nodes, for nodes with multiple out-going links, we randomly retain one. For these nodes, their $l_u=1$. Mathematically, we have obtained a forest of trees, where the root of each tree is a local leader \cite{blondel2008local}, and is also a potential community center. For most nodes, except local leaders, this process identifies a local hierarchy, with an asymmetric leader-follower relation. Third, to identify the upper level for local leaders along the hierarchy, LS uses a local breadth-first search starting from each local leader $u$ and stop the search when encountering a nearest local leader $v$ with $k_v\geq k_u$ and assign the shortest path length on the original network $d_{uv}$ to $l_u$, which will be no smaller than two, i.e., $l_u\geq 2$. LS uses the product of rescaled degree $k_u$ and rescaled distance $l_u$ to quantitatively to identify community centers. Finally, community labels are then assigned along the reverse direction of the previously detected local hierarchy from community centers. 

Community centers can also be selected in a posterior way, for example, after obtaining the community partition by Louvain algorithm \cite{blondel2008fast} or other community detection algorithms, we can then choose the most central one in each community. Here, it is chosen according to coreness, if there are multiple nodes with the same coreness, we then choose the one with the highest degree. We also tried other centrality measures, e.g., first based on h-index \cite{lu2016h} and then degree, the chosen centers are almost the same. Such selection criteria or sequence can be easily tailored to accommodate different perspectives regarding node importance.

\subsection*{Normalized compression length of a network} 
To quantify the information encoded in networks constructed with different $\varepsilon$, we employ the normalized compression length. It is based on the principle of lossless compression in information theory and adheres to the ``minimum description length'' criterion: the more regular the network structure, i.e., the more predictable the inter-node connection patterns, the shorter the shortest description length of its binary encoding, and the smaller the normalized value \cite{Choi2012Compression,sun2020revealing}. 

The minimum achievable compression length $L$ of a network can be derived using a lossless compression approach \cite{Choi2012Compression}. 
This method has been validated as an optimal solution for compressing random networks and also demonstrates high efficiency when applied to a wide range of real-world networks. 
Given that isomorphic networks share identical structural properties, which only differs in node labeling, the compression process begins by translating the network's structural information into a binary code string. Notably, this binary string retains sufficient information to reconstruct any network isomorphic to the original one. 
To further eliminate inherent correlations within the network structure, the binary code string undergoes a second compression step using a recursive arithmetic encoding technique \cite{skretting1999improved}. This encoder is specifically designed to leverage interdependencies between symbols in the binary string, enhancing compression efficiency. 
Following these two sequential compression steps, the length of the resulting bit string aligns closely with the network's structural entropy \cite{Choi2012Compression,shannon1948entropy,shannon1953lattice}. This alignment enables the bit string length to serve as a reliable metric for quantifying the network's randomness, consistent with the principles outlined in Shannon's source coding theorem \cite{cover1999elements}. 

To eliminate the influence of network size, i.e., the number of nodes $N$ and the number of edges $E$, a random network generated based on the Erd\"os–R\'enyi model \cite{bollobas2011random,Choi2012Compression} is introduced as a benchmark. Such a random network has the same number of nodes and edges as the target network, and its  length $\mathcal{R}$ is calculated. 
The final normalized compression length $L^{*}$ is defined as: $L^{*} = L /\mathcal{R},$
where $\mathcal{R}=\binom{N}{2} h(q)-N \log N$, $q=E/{\tiny \begin{pmatrix} N \\ 2 \end{pmatrix}}$ is the node pair connection probability, and $h(q)=-q\log q-(1-q)\log(1-q)$ is the binary entropy (with logarithm base 2). A smaller $L^{*}$ indicates that the target network has a higher structural regularity and information encoded in the network compared to the random null model, which is hardest to compress due to its complete randomness.

\subsection*{Data availability}
All network and vector datasets needed to evaluate the conclusions in the paper are publicly available and present in the paper and/or the Supplementary Materials.  

\subsection*{Code availability}
Computer code for our RCC algorithm and code implementing the analysis described in this paper and other information is online at \url{https://github.com/UrbanNet-Lab/RCC-Regression_by_Community_Centers}.  

\subsection*{Author contributions statement}
All authors conceived the project. R.Li and T.B. wrote the analysis code, performed the numerical simulations and data analyses, and developed and implemented the methodology. R.L. drafted the manuscript with input from all authors. All authors reviewed and edited the manuscript. 

\subsection*{Competing Interest declaration}
The authors declare no competing interests.

\section*{Acknowledgments}
This work receives financial supports from the National Natural Science Foundation of China (Grant No. 72371014, 72225012, 72288101, and 71822101), and supported by Shanghai Key Laboratory of Intelligent Information Processing, Fudan University (Grant No. IIPL-2025-RD1-05). R.Lambiotte acknowledges support from the EPSRC Grants EP/V013068/1, EP/V03474X/1, and EP/Y028872/1.

\bibliography{sample}

\end{document}